\documentclass[sigconf,natbib=true,anonymous=false,dvipsnames, svgnames]{acmart}
\usepackage{adjustbox}
\usepackage{multirow} 
\AtBeginDocument{%
  \providecommand\BibTeX{{%
    \normalfont B\kern-0.5em{\scshape i\kern-0.25em b}\kern-0.8em\TeX}}}

\copyrightyear{2023}
\acmYear{2023}
\setcopyright{acmlicensed}\acmConference[SIGIR '23]{Proceedings of the 46th International ACM SIGIR Conference on Research and Development in Information Retrieval}{July 23--27, 2023}{Taipei, Taiwan}
\acmBooktitle{Proceedings of the 46th International ACM SIGIR Conference on Research and Development in Information Retrieval (SIGIR '23), July 23--27, 2023, Taipei, Taiwan}
\acmPrice{15.00}
\acmDOI{10.1145/3539618.3592071}
\acmISBN{978-1-4503-9408-6/23/07}

\pdfoutput=1

\begin{document}

\title{The tale of two MS MARCO - and their unfair comparisons}
% Where is my ti

\author{Carlos Lassance}
\email{carlos.lassance@naverlabs.com}
\orcid{0000-0002-7754-6656}
%\email{webmaster@marysville-ohio.com}
\affiliation{%
  \institution{Naver Labs Europe}
  \city{Meylan}
  \country{France}
}

\author{Stéphane Clinchant}
\email{stephane.clinchant@naverlabs.com}
\orcid{0000-0003-2367-8837}
%\email{webmaster@marysville-ohio.com}
\affiliation{%
  \institution{Naver Labs Europe}
  \streetaddress{6 chemin de Maupertuis}
  \city{Meylan}
    \country{France}
}
%\renewcommand{\shortauthors}{Trovato, et al.}

\begin{abstract}
The MS MARCO-passage dataset has been the main large-scale dataset open to the IR community and it has fostered successfully the development of novel neural retrieval models over the years. But, it turns out that two different corpora of MS MARCO are used in the literature, the official one and a second one where passages were augmented with titles, mostly due to the introduction of the Tevatron code base. However, the addition of titles actually leaks relevance information, while breaking the original guidelines of the MS MARCO-passage dataset. In this work, we investigate the differences between the two corpora and demonstrate empirically that they make a significant difference when evaluating a new method. In other words, we show that if a paper does not properly report which version is used, reproducing fairly its results is basically impossible. Furthermore, given the current status of reviewing, where monitoring state-of-the-art results is of great importance, having two different versions of a dataset is a large problem. This is why this paper aims to report the importance of this issue so that researchers can be made aware of this problem and appropriately report their results.\end{abstract}

%Recently neural retrievers have been shown as capable alternatives to term-based strategies for in-domain retrieval. Research on those models has been mostly developed thanks to the open-sourcing of data and models, focusing mostly on the MS MARCO passage dataset. However, in the middle of the way, two different corpora of MS MARCO started being used, the base one and a second one with added titles, mostly due to the introduction of the Tevatron framework. However, the addition of titles to the passages potentially leaks label information and actually breaks the original guidelines of MS MARCO. In this work, we investigate the differences between the two corpora and demonstrate empirically that they make a great difference when evaluating a method. In other words, we show that if a paper does not properly report which version it is using, reproducing and fairly comparing to it is basically impossible.\end{abstract}




\maketitle

\section{Introduction}
Benchmarking and reproducibility are important activities in any scientific discipline but are especially well recognized in the information retrieval community. Not only the TREC challenges pioneered a systematic and rigorous evaluation of search systems, but the main IR conferences SIGIR and ECIR have also dedicated reproducibility tracks for several years now.

Thus, research on neural models has been mostly fostered thanks to the open-sourcing of data and models, which allow for an easier path to both reproducibility and benchmarking.
With the deep learning revolution, neural models were in need of large scale data and
the MS MARCO-passage dataset has played a leading role in the development of novel neural models, being similar to an ImageNet dataset for IR. 
%Recently neural retrievers have been shown as an interesting avenue~\cite{} to replace term-based retrieval strategies such as BM25~\cite{}. 
MS MARCO-passage~\cite{msmarco}, builds upon data from a real search engine, Bing, and allowed training using vast amounts of data for training (500k annotated training queries) and also evaluation~\footnote{~100k dev queries, even if most of the time only 6980 are used}.

For model development, both GitHub~\cite{github} and HugginFace transformers~\cite{huggingface} have enabled either to share  training code or models more easily than before. Therefore, one could more easily benchmark against other models or even build on top of them without huge reproducibility costs. Furthermore, efforts in making replicable results improve the impact of a research paper as noted in~\cite{lin2022building,raff2022does}, even if just code by itself may not be enough to make a paper reproducible~\cite{raff2022siren}. Regarding neural IR, one of the most notable case of sharing training code and making dense retrievers replicable is the Tevatron framework~\cite{tevatron} which was first used in~\cite{cocondenser} and has become almost a popular framework for training neural retrievers. However, the Tevatron repository relies on a ``titled'' MS MARCO-passage corpus~\footnote{\url{https://huggingface.co/datasets/Tevatron/msmarco-passage-corpus}} which creates an unfair comparison the ``official'' MS MARCO-passage corpus. 

The goal of this paper is to warn the IR community with this simple statement which has a huge impact: \emph{there are actually \textbf{two versions} of the \textbf{MS MARCO-passage} dataset used in papers}: 1) the official one and 2) a title augmented version. To our knowledge, the MSMARCO passage titled version was introduced in~\cite{qu2020rocketqa}, which has more than 200 citations by the time of submission, but gives no details on the use and from where titles were collected. We will demonstrate that the titles do make a difference with respect to effectiveness, which then makes the comparison to other models impossible. Overall, our main contributions are:
\begin{enumerate}
    \item investigate where the MS MARCO titled version came from and how it differs from the original MS MARCO;
    \item identify the biases generated by the modified corpus;
    \item evaluate how models behave due to these different versions;
    \item show that the gains obtained by the different MS MARCO version do not generalize to other dataset.
\end{enumerate}

Before going into the motivation, we would like to acknowledge again the efforts of the creators of MS MARCO and Tevatron for helping the community and making all this research easier. We do not want to blame any framework, nor datasets, nor the researchers that used any version of them, but mostly to bring to light this problem created by using an ``unnoficial MS MARCO''. As of now, it simply is difficult to reproduce and compare different approaches, and more so, fairly evaluating papers during review time. This paper is structured as follows: in Section~\ref{sec:motivation}, we discuss the MS MARCO dataset in detail and its ``titled'' version. In Section~\ref{sec:experiments}, we proceed to experiments to assess the impact of the title before concluding.

%\section{Motivation}
\section{Datasets}
\label{sec:motivation}

In this section, we motivate our study, by first presenting the MS MARCO dataset, what is its modified corpus, and what biases we can detect just by analyzing the modified corpus and the queries.

\subsection{The MS MARCO passage dataset}
MS MARCO (MicroSoft MAchine Reading COmprehension)~\cite{msmarco} is a large-scale dataset focused on machine reading comprehension. It is a large-scale dataset aiming at reproducing real-world scenarios in Web search with
little or noisy annotations based on click logs. One specificity of this dataset is to have only a single relevant item for a given query. Since its release, MS MARCO has been used primarily in IR evaluation in the TREC Deep Learning track~\cite{trec19,trec20} but also in NLP evaluation such as keyphrase extraction, and conversational search. 
The MS MARCO passage dataset consits of 8,8M  passages and 1M queries with roughly 400M of training triplets. %An important note from the dataset website is one shall note  extract additional data from MS MARCO question answering task 
%\footnote{\url{https://microsoft.github.io/MS MARCO/Datasets.html}}.
%\begin{quote}
%IMPORTANT NOTE: It is prohibited to use evidence from the MS-MARCO Question Answering task in your submission. That dataset reveals some minor details of how the MS MARCO dataset was constructed that would not be available in a real-world search engine; hence, should be avoided.
%\end{quote}
The TREC Deep Learning track organizers, which include most of the original authors of MS MARCO, not only interdict the use of passage-document mapping~\cite{trec21} as shown in the cited text: 
\begin{quote}
The v1 data had several problems. The corpus was generated based on the queries, such that each passage and each document is in the corpus due to one of our million original queries. For each document in the corpus there may only be one passage in the passage dataset (and on average 2.8 passages per document), but that passage was identified by
Bing in relation to one of the MS MARCO queries, possibly a test query. This is unrealistic, since a real system would be able to generate many candidate passages per document, and would not know what the test queries will be ahead of time. Therefore, we had to forbid participants from considering the passage-document mapping. \end{quote} In their guidelines~\footnote{\url{https://github.com/microsoft/msmarco/blob/42417041b22ef5fa80bbe150212c9f55c999e820/Datasets.md}}, the authors also discouraged adding data to the MS MARCO-passage corpus.

%\begin{quote}
%IMPORTANT NOTE: You are allowed to use external information while developing your runs. However, it is prohibited to use any datasets from MS MARCO.org in your submission except those listed above. The original MS MARCO question-answering dataset reveals minor details of how the dataset was constructed that would not be available in a real-world search engine; hence, should be avoided.
%\end{quote}

\subsection{MS MARCO ``titled'' corpus and Tevatron}
Thanks to MS MARCO, and with the research on dense retrieval and the effort to share and reproduce results, several GitHub code bases have been developed to promote reusability and easier research. Tevatron is one of them and aims at being a simple and efficient toolkit for training and running dense retrievers with pre-trained language models. %The toolkit is thought to have a modularized design for easy research; a set of command line tools are also provided for fast development and testing and  a set of easy-to-use interfaces to Huggingface's state-of-the-art pre-trained transformers.
Overall, Tevatron is an easy-to-use library to do research on dense retrievers, question answering  and large-scale information retrieval. However, tutorials and papers using the framework use the MS MARCO ``titled corpus'' leading to a  reproducibility problem. Indeed, trying to reproduce models trained with MS MARCO-passage have already detected problems in doing so, for instance, issue 66 \footnote{ \url{https://github.com/texttron/tevatron/issues/66}}  observes some discrepancy between the datasets and issue 34 \footnote{\url{https://github.com/texttron/tevatron/issues/34}} talks about problems when trying to reproduce the results until they discover that the problem comes from the MS MARCO ``titled'' corpus.

Finally, we note that ``titled'' is not the only non standard MSMARCO passage on huggingface. For example, \url{https://huggingface.co/datasets/ms_marco} considers v1.1 as the original Q\&A MS MARCO and v2.1 as the MS MARCO-passage dataset. However, there is also the MS MARCO v2 dataset~\cite{trec21}, which is an expansion of the MS MARCO-passage, thus it is an expansion of what was previously called v2.1, making it confusing to newcomers. In this work, we consider only the MS MARCO-passage dataset and its ``titled'' version.

\subsection{Examples from the ``titled'' corpus}

Let's start by looking at two examples. First, let us consider the following passage with id 4216689: 
\begin{quote}
The project charter is a single, consolidated source of information about the project in terms of initiation and planning. Basically, the project charter defines the boundaries of the project, no matter what type of project management methodology you are using. It is much more than an effective planning tool.
\end{quote} This passage is considered a positive for query 787255 ``what is project charter in project management'', note how an exact match system would find most of the query words on the passage, safe for ``what''. However, if we add the title ``What Is a Project Charter?'', it makes for a much easier search, as all the query words are now present in the text, and actually just looking at the title we have almost an exact match. Another example is passage 7799738: 
\begin{quote}
The Maastricht Treaty (formally, the Treaty on European Union or TEU) undertaken to integrate Europe was signed on 7 February 1992 by the members of the European Community in Maastricht, Netherlands. On 9–10 December 1991, the same city hosted the European Council which drafted the treaty.
\end{quote} This passage is considered a positive for query 928478 ``what year was the maastricht treaty''. Looking into the MS MARCO-passage corpus, there are 101 passages containing the words Maastricht and treaty, but if we look into the titled corpus, only 8 passages have Maastricht treaty on the title (including the positive 7799738), making for a much easier search. Actually, looking into the 8 passages with that title, 5 of them could be considered correct for that query, which also shows a bit of the bias of the MS MARCO-passage dataset as the sparse labels make for various false negatives.

%rep -n -i 7188777 raw.tsv
%7188778:7188777	Human Papillomavirus Vaccine Cost [SEP] For patients not covered by health insurance, the cost of HPV vaccination typically includes: shot administration fees and the cost of the three required doses of the vaccine at about $125 each, for a total of $400 to $500. 1  Many health insurance plans cover HPV vaccination, but only for females in the recommended age group.
%1047794 cost of hpv vaccine cdc

%For instance, for the query '928478	what year was the maastricht treaty'.
%The relevant doc is this one:
%7799738 Maastricht Treaty [SEP] The Maastricht Treaty (formally, the Treaty on European Union or TEU) undertaken to integrate Europe was signed on 7 February 1992 by the members of the European Community in Maastricht, Netherlands. On 9–10 December 1991, the same city hosted the European Council which drafted the treaty.

%There are 104 documents containing the exact word maastricht and treaty. There are actually 8 passages with tile 'Maastricht Treaty', hence reinforcing exact match signals.

\subsection{Some statistics about the ``titled'' corpus}

To start, let us look at how many passages have an added title in this corpus. From the entire 8.8M passages, only 5.7M (64.5\%) have been associated with a title. Our analysis shows that while document->passage is an injective function, the passage->document is not injective, in other words, the same passage may be found in more than one document, making it impossible to perform title association a-posteriori. Considering that the original labels come from the document set and that each query is associated with a document, not being a repeated passage is a strong signal that it may be a positive: on the other hand, if a passage belonging to multiple documents is considered a positive this means that we are missing a relevance label for the other documents containing the passage.

Now let us look into the 6980 dev set of MS MARCO, which is where most papers evaluate themselves, using MRR@10. Considering that MRR@10 is the measure they look into, we care about how many queries have at least one positive with a title, as MRR@10 only cares about the position of the ``best'' positive. Actually, 73.4\% of the queries, have a positive with an added title, meaning that passages with titles have $1.5\times$ more probability of being a positive compared to a passage without title (64.5\%$\rightarrow$73.4\% vs 35.5\%$\rightarrow$28.8\%), even if we discount the fact that there are more titled than untitled passages. In other words, given the sparse labels of the MS MARCO dataset, an untitled passage has more of a chance to be a false negative than a passage with a title, which means that knowing that a passage has a title is a potential leak of label information.

We further hypothesize the possible leak effect by looking into the TREC-DL 2019 and 2020 query sets, which do not use title information when creating their relevance labels. The TREC-DL datasets are evaluated using nDCG@10 meaning that \textit{at most} we deal with 10 positives, and labels are assigned in a 0 to 3 relevance level without looking into titles, with labels 2 and 3 being considered positive and label 1 just relevant but not necessarily positive (disregarded for recall@k metrics). We still see the bias on the TREC-DL 19 where 73.2\% of the positives have titles, while only 67\% of the negatives have titles. However considering that there's an average of 15 positives without titles per query, this bias towards ``titled passages'' should not impact the nDCG@10 evaluation. On the other hand, looking at TREC-DL 20 we still see an over-representation of positive titled passages (67\% which is higher than the 64.5\% of passages with titles), the over-representation is smaller and is actually mostly the same for both positives and negatives (67\% positive and 66\% negative have titled passages)

\subsection{Motivation of this work}

To summarize, our main motivation is that the ``MS MARCO'' titled corpus is not exactly the same as the one considered the default MS MARCO-passage corpus. This ``titled'' version is biased toward positive documents in the dev set and breaks the guidelines from the TREC-DL, which is organized by the authors of MS MARCO. If the authors do not explicitly discuss which version they are using, as it is done in~\cite{cocondenser}, it may lead to making the methods impossible to reproduce, even if the authors share their trained model. More so, if they compare against works that did not use this modified corpus, such as TAS-B~\cite{tasb}, ColBERTv2~\cite{colbertv2} and SPLADE++~\cite{spladev2}, they do so in an unfair fashion. Considering the biases we detailed in the previous subsection we would expect models trained and evaluated with titles to have better effectiveness in the MS MARCO dev, but that it should not appear in TREC-DL 19 and TREC-DL 20, meaning that it does not generalize.

\section{Experiments}
\label{sec:experiments}

Now that we have introduced the ``titled'' version of MS MARCO, we now look into the problems it causes empirically. We look into three types of models: i) dense retrievers; ii) sparse learned retrievers and iii) cross-encoding rerankers. Note that this could be extended to other models, but we chose the three that we had most familiarity with. In more detail, we use DPR-cls~\cite{dpr} as our dense retriever, SPLADE-max~\cite{spladev2} as our learned sparse retriever and RankT5-3b encoder only~\cite{rankt5}, monoElectra-large~\cite{monoelectra} as our cross encoding rerankers. We train models on either the ``titled'' or the official MS MARCO-passage dataset. Titled passages are generated using ``Title [SEP] Passage''. We use only contrastive training (i.e. no distillation) and hard-negatives\footnote{ \url{https://huggingface.co/datasets/sentence-transformers/msmarco-hard-negatives}} to train the first stage models and the negatives from SPLADE to train the reranker. First-stage retrieval training uses 8 queries per batch, and 32 negatives per query, and trains for 3 epochs. We use DistilBERT~\cite{distilbert} as the pre-trained language model for the first stage models and for reranking Electra-large and T5-3B. Statistical significance is computed only when we directly compare the original vs ``title'' datasets (i.e. only Tables 1 and 4 and only on the same model with different corpus), with Student's t-test and $p\leq 0.05$.

\subsection{First-Stage Retrievers}
%\subsection{Impact on first stage retrievers?}

We first test the impact of the ``titled'' corpus on first stage retrievers. We train the models using the same training scheme, changing only the corpus that is used. Evaluation on the BEIR dataset differs as the methods trained without title are evaluated without any separator between the BEIR title and passage, while methods trained with title use the [SEP] separator. We present the results in Table~\ref{tab:first_stage}. As expected, the models trained and evaluated with the ``titled'' corpus had an increase in effectiveness in the MS MARCO-passage dev dataset, that was not translated to the TREC-DL query sets, which not only shows that the same training with the different corpora leads to unfair comparisons on the MS MARCO-passage dev dataset, but that the improvement does not generalize to query sets from the same dataset that do not have the label leakage from the titles. To give an idea of the magnitude of the improvement due to the titled corpus, consider the DPR-CLS which `` improves'' from 33.8 to 35.3, a gain of 1.5, which is equivalent to the improvement from the initial distillation models from~\cite{viennadistil} and the TAS-B model from~\cite{tasb} (32.6 to 34.0). Finally, when looking at zero-shot effectiveness on the BEIR dataset, we actually find a slight improvement when training with titles on the SPLADE setting, which was unexpected. We will delve into this in the next section.

\begin{table}[ht]
\caption{First stage retrieval comparison between models trained and evaluated with and without the ``titled corpus''. BEIR is the average over the 13 easily available datasets. $^\dagger$ represents statistically significant difference.}
\label{tab:first_stage}
\adjustbox{max width=\columnwidth}{%
\begin{tabular}{c|cc|cc|c}
\toprule
\multirow{2}{*}{Corpus} & \multicolumn{2}{c}{MSMARCO} & TREC-19 & TREC-20 & BEIR \\
                      & MRR@10 & R@1k   & nDCG@10 & nDCG@10 & nDCG@10 \\
\midrule
\multicolumn{6}{c}{\textbf{DPR-CLS models (dense retrieval)}}                                    \\
\midrule
 original                     & 33.8   & 95.4\% & 65.8    & 67.1    & 39.4    \\
with title                     & 35.3   & 96.1\% & 66.0    & 63.1    & 38.2    \\
\midrule
title -  original & 1.5$^\dagger$    & 0.6$^\dagger$    & 0.2     & -4.0$^\dagger$    & -1.2    \\
\midrule
\multicolumn{6}{c}{\textbf{SPLADE models (sparse retrieval)}}                                     \\
\midrule
 original                     & 37.7   & 97.3\% & 72.3    & 70.5    & 44.8    \\
with title                     & 38.5   & 97.7\% & 70.6    & 67.1    & 45.3    \\
\midrule
title -  original  & 0.8$^\dagger$    & 0.4$^\dagger$    & -1.6    & -3.5    & 0.6    \\
\bottomrule
\end{tabular}
}
\end{table}

\subsubsection{Variants on how to use the title information}

To better understand which parts of the ``titled'' dataset is actually useful, we create 4 versions of the dataset, with increased information: i) (DEFAULT) MS MARCO-passage corpus without changes; ii) (SEP) adding [SEP] at the start of all passages; iii) (INFO) adding the information that a passage has a title, but not the title itself, i.e. we add ``TITLE [SEP]'' at the beginning of titled passages and just [SEP] to the others; and iv) (``titled'') the ``titled'' corpus. We first train and evaluate the SPLADE model with each of these corpuses and present the results in Table~\ref{tab:ablation_1}. Overall, using the title content is the best on MS MARCO. However, on BEIR, it does not make a difference as the model may have simply learn to use the [SEP] to match titles in BEIR: the BEIR improvement from the ``titled'' corpus actually disappears when the models are trained with the [SEP] token, and that the decrease in effectiveness on the TREC datasets may come from overfitting on the title use, and not just the information that a passage has a title.

\begin{table}[ht]
\caption{First stage retrieval with SPLADE under variants on how to use the title information.}
\label{tab:ablation_1}
\adjustbox{max width=\columnwidth}{%
\begin{tabular}{c|cc|cc|c}
\toprule
\multirow{2}{*}{Corpus} & \multicolumn{2}{c}{MSMARCO} & TREC-19 & TREC-20 & BEIR \\
                         & MRR@10 & R@1k   & nDCG@10 & nDCG@10 & nDCG@10 \\
\midrule
passage (default)      & 37.7   & 97.3\% & 72.3    & 70.5    & 44.8    \\
 {[SEP]} + passage (SEP)             & 37.9   & 97.4\% & 72.6    & 70.5    & 45.7    \\
INFO + [SEP] + passage (INFO)  & 37.6   & 97.3\% & 72.9    & 69.5    & 45.7    \\
Title + [SEP] + passage (``titled'') & 38.5   & 97.7\% & 70.6    & 67.1    & 45.3   \\
\bottomrule
\end{tabular}
}
\end{table}

\subsubsection{Changing the corpus from training to evaluation} 

We now take the 4 models we trained for the previous experiment and evaluate them over the 4 different corpora we developed. Results for MS MARCO dev are presented in Table~\ref{tab:ablation_2}, where the diagonal means that a model is trained and tested with the same corpus. We first note that adding SEP at the start of passages can actually improve the default model, as it is the only off-diagonal bolded value (i.e. best value in the column is outside the main diagonal). We also note how the titles actually reduce the effectiveness of models unless the model was trained with it, when it actually outperforms all the other methods, meaning that both a) using the titles has to be learned and b) training with titles actually overfits to them as testing without them makes for worse results.

\begin{table}[ht]
\caption{SPLADE  under variants on how to use the title information, using different combinations of train/evaluation corpus.
%INFO refers to adding the information that a passage has a title, while SEP refers to adding [SEP] to the start of all passages.
\small{Bolded is best of training corpus (i.e. best in column), and underlined is best of test corpus (i.e. best in row).}}
\label{tab:ablation_2}
\adjustbox{max width=\columnwidth}{%
\begin{tabular}{ccccc}
\toprule
\multirow{2}{*}{Test corpus} & \multicolumn{4}{c}{Training corpus}                                       \\
                             & default & SEP & INFO & ``titled'' \\
\midrule
MSMARCO-passage (default)             & \underline{37.7}            & 37.2        & 37.1         & 36.8                       \\
{[SEP]} + passage (SEP)                 & \textbf{37.9}            & \textbf{\underline{37.9}}        & 37.5         & 37.6                       \\
INFO + [SEP] + passage (INFO)                & 37.4            & 37.2        & \textbf{\underline{37.6}}         & 37.5                       \\
TITLE + [SEP] + passage (``titled'')   & 36.9            & 37.0        & 36.8         & \textbf{\underline{38.5}}    \\                  
\bottomrule
\end{tabular}
}
\end{table}

\subsection{Cross-encoding rerankers}
Finally, we now look into how the titles impact cross-encoding rerankers. Cross-encoding rerankers use both the query and the passage as input to get the score, instead of relying at the separate representations, thus leading to better effectiveness in general. We use an ensemble of ColBERTv2, SPLADE and DocT5 (so 3 models that do not use titles) as the first stage in order to generate an ensemble of candidates that i) does not use the ``titled'' version; and ii) is as close to a state of the art first stage as possible) and present the results in Table~\ref{tab:reranker}. As expected, results with rerankers also follow the same tendency as first stage rankers: adding titles improves effectiveness in the MS MARCO dev set making it impossible to compare fairly, while not reproducing that advantage in the TREC-19 and TREC-20 query sets.

\begin{table}[ht]
\caption{Cross-encoding Reranking Comparison. \small{MSMARCO reports MRR@10. TREC-19 and TREC-20 report nDCG@10. TOP10 and 50. refers to the amount of reranked candidates. $^\dagger$ represents a statistically significant difference.}}
\label{tab:reranker}
\adjustbox{max width=\columnwidth}{%
\begin{tabular}{c|cc|cc|cc}
\toprule
\multirow{2}{*}{corpus} & \multicolumn{2}{c|}{MSMARCO} & \multicolumn{2}{c|}{TREC-19} & \multicolumn{2}{c}{TREC-20} \\
                      & TOP10 & TOP50 & TOP10 & TOP50 & TOP10 & TOP50 \\
\midrule
\multicolumn{7}{c}{Baseline (First   Stage)}                          \\
\midrule
original                          & \multicolumn{2}{c|}{39.3}    & \multicolumn{2}{c|}{76.3}    & \multicolumn{2}{c}{75.3}    \\
\midrule
\multicolumn{7}{c}{MonoElectra-Large}                                 \\
\midrule
 original                &    40.3 & 39.5 & 78.4 & 76.0 & 78.5 & 76.7 \\
with title               &    41.2 & 40.6 & 78.4 & 74.7 & 77.6 & 73.9 \\
\midrule
title -   original       & 0.9$^\dagger$ & 1.1$^\dagger$ & 0.0 & -1.2 & -1.0 & -2.7$^\dagger$  \\
\midrule
\multicolumn{7}{c}{RankT5-3b encoder   only}                          \\
\midrule
 original                     & 43.0  & 42.9  & 78.4  & 76.8  & 78.3  & 77.0  \\
with title                     & 44.2  & 44.3  & 78.3  & 76.3  & 78.4  & 75.7  \\
\midrule
title -   original & 1.2$^\dagger$   & 1.4$^\dagger$   & -0.1  & -0.5  & 0.1   & -1.3 \\
\bottomrule
\end{tabular}
}
\end{table}

\section{Conclusion}

We have shown empirically that the ``titled'' MS MARCO passage corpus not only breaks the guidelines of the dataset, but also that models trained using that version of the corpus have an unfair advantage. As of the SIGIR23 short paper submission, the Tevatron arxiv paper~\cite{tevatron} has 21 citations mostly from IR researchers, and most of the time without adding the caveat that they are using a non-standard MS MARCO dataset. Most of those papers will then compare their results against other models such as TAS-B~\cite{tasb}, ColBERTv2~\cite{colbertv2} or SPLADE++~\cite{spladev2} that do not use the modified version of the corpus, which leads to a false comparison. We hope that by raising the alert with this work, future research actually separates models trained/evaluated with the different versions of the MS MARCO dataset, leading to more reproducible research and research that performs fair comparisons between the different models. We want to finish this paper by reiterating that we do not aim to blame any work referred here, but simply to state the problems that arose from the simple decision of not using the standard MS MARCO-passage dataset, even more so as most researchers were probably not aware of that decision.

\bibliographystyle{ACM-Reference-Format}
\bibliography{sample-base}

\end{document}